\documentclass[aps, twocolumn, pre, floatfix, superscriptaddress]{revtex4-2}

\usepackage{graphicx}
\usepackage{bm}
\usepackage{amsmath,amssymb,amsfonts,amsthm}
\usepackage{physics}
\usepackage{color}

\usepackage{ulem}

\definecolor{darkblue}{rgb}{0,0,0.65}
\definecolor{darkgreen}{rgb}{0.0,0.6,0.0}
\definecolor{darkred}{rgb}{0.6,0.0,0.0}
\definecolor{cyan1}{rgb}{0.0, 0.6, 0.6}
\usepackage[%
  pdfstartview=FitH,%
  breaklinks=true,%
  bookmarks=true,%
  colorlinks=true,%
  anchorcolor=black,%
  citecolor=blue,
  filecolor=black,%
  menucolor=black,%
  urlcolor=darkred,%
  linkcolor=darkgreen,%
 ]{hyperref}

\providecommand{\ket}[1]{\ensuremath{\left|{#1}\right.\rangle}}

\providecommand{\pd}[2]{\ensuremath{\frac{\partial#1}{\partial#2}}}
\providecommand{\pds}[2]{\ensuremath{\frac{\partial^2#1}{\partial#2^2}}}
\providecommand{\dE}{\ensuremath{\Delta E}}

\graphicspath{{.}{./Figures/}}

\begin{document}
	
\title{Entropy and Temperature in finite isolated quantum systems}

\author{Phillip C. Burke}

\affiliation{Department of Theoretical Physics, Maynooth University, Maynooth, Kildare, Ireland}

\author{Masudul Haque}

\affiliation{Institut f\"ur Theoretische Physik, Technische Universit\"at Dresden, 01062 Dresden, Germany}

\affiliation{Department of Theoretical Physics, Maynooth University, Maynooth, Kildare, Ireland}

\affiliation{Max-Planck Institute for the Physics of Complex Systems, Dresden, Germany}

\date{\today}

\begin{abstract}
	We investigate how the temperature calculated from the microcanonical entropy compares with 
	the canonical temperature for finite isolated quantum systems.  We concentrate on systems 
	with sizes that make them accessible to numerical exact diagonalization.  We thus characterize 
	the deviations from ensemble equivalence at finite sizes.  We describe multiple ways to compute 
	the microcanonical entropy and present numerical results for the entropy and temperature computed 
	in these various ways.  We show that using an energy window whose width has a particular energy 
	dependence results in a temperature with minimal deviations from the canonical temperature.
\end{abstract}

\maketitle

\section{Introduction}

In recent years, there has been considerable interest in understanding how statistical mechanics
emerges from the quantum dynamics of isolated many-body systems.  Such considerations invariably
require a correspondence between energy, a quantity well-defined in quantum mechanics, and
temperature, which is necessary for a statistical-mechanical description.  Since temperature is not
a priori defined in quantum mechanics, assigning temperatures to energies is a nontrivial issue.

Ideas regarding thermalization in isolated quantum systems, e.g., the eigenstate thermalization
hypothesis (ETH) \cite{Deutsch_PRA1991, Srednicki_PRE1994, Srednicki_1996, Srednicki_1999,
  Rigol_thermalization_Nat2008, AlessioRigol_Chaos_AdvPhys2016,
  Reimann_NJP2015,Deutsch_RepProgPhys2018, Mori_Ikeda_Ueda_thermalizationreview_JPB2018} and its
extensions or variants are often tested or verified using numerical ``exact diagonalization''
calculations \cite{ Rigol_thermalization_Nat2008, Rigol_Quenches_PRA2009,
  Rigol_thermalization_PRL2009, BiroliKollathLauchli_FluctationsAndThermalization_PRL2010,
  SantosRigol_Thermalization_PRE2010, SantosRigol_LocalizationAndThermalization_PRE2010,
  RigolSantos_Chaos_PRA10, Roux_quantumquenches_PRA2010, NeunhahnMarquardt_thermalization_PRE2012,
  RigolSrednicki_Thermalization_2012, SantosPolkovnikovRigol_typicality_PRE2012,
  SantosPolkovnikovRigol_typicality_PRE2012, GenwayHoLee_ThermalizationSmallHubbard_PRA2012,
  Steinigeweg_etal_EthThermalization_PRE2013, KimIkedaHuse_TestETH_PRE2014,
  BeugelingHaque_ThermalizationScaling_PRE2014, SorgVidmarHeidrichMeisner_thermalization_PRA14,
  Steinigeweg_etal_ETHLimits_PRL2014, FratusSrednicki_EstateThermalization_PRE2015,
  BeugelingMoessnerHaque_matrixelements_PRE2015, NandkishoreHuse_MBL_AnnRev2015,
  JohriNandkishoreBhatt_MBL_PRL2015, Luck_SmallSystem_JoPA2016,  
  MondainiFratusSrednickiRigol_ETH2D_PRE2016,
  ChandranSchulzBurnell_AnyonETH_PRB2016, LuitzBarLev_AnomalousThermalization_PRL2016,
  AlessioRigol_Chaos_AdvPhys2016, MondainiRigol_ETH2DII_PRE2017, Dymarsky_Lashkari_Liu_PRE2018,
  GarrisonGrover_SingleEstate_PRX2018, YoshizawaIyodaSagawa_NumericalETH_PRL2018,
  KhaymovichHaqueMcClarty_Behemoths_PRL2019, Vidmar_HeidrichMeisner_ETHpolaron_PRB2019,
  MierzejewskiVidmar_ETH_PRL2020, BrenesPappalardiGoold_EntanglementETH_PRL2020,
  Santos_SpeckOfChaos_PRR2020, LeBlondRigol_EstateThermalization_PRE2020,
  BrenesRigol_etal_ETHPerturbed_PRL2020, Noh_ETH_PRE2021,
  Sugimoto_Hamazaki_Ueda_ETH_localRMT_PRL2021, SchonleJansenRMeisnerVidmar_ETHAuto_PRB2021,
  Nakerst_Haque_ETHclassicallimit_PRE2021, Richter_Steinigeweg_Dymarsky_Gemmer_PRL2022,
  SugimotoHamazakiUeda_ETHLongRange_PRL2022}.  As a result, quantum systems with Hilbert space
dimensions between $\sim10^3$ and $\sim10^5$ have acquired particular relevance.  It is therefore
important to ask how meaningful various definitions of thermodynamic quantities like temperature or
entropy are for finite systems, particularly systems of sizes typically treated by full numerical
diagonalization.  In this work, we critically examine different ways of calculating entropy from the
energy eigenvalues of finite systems and compare the temperature derived from the entropy with the
so-called `canonical' temperature.

The most common definition of temperature for finite isolated quantum systems is the canonical
temperature.  This is obtained for any energy  $E$ by inverting the canonical equation 
\begin{equation}\label{eq:canonical_energy}
	E = \langle H \rangle \ = \ \frac{\tr( e^{-\beta H}H)}{\tr( e^{-\beta H})} 
	\ = \ \frac{ \sum_{n} e^{-\beta E_{n}}E_{n} }{\sum_{n} e^{-\beta E_{n}}}. 
\end{equation}
If the eigenvalues $\{E_n\}$ of a system are known, then this relationship provides a map
between energy and the canonical temperature  $T_c=(k_B\beta_{c})^{-1}$.  (Here $k_B$ is the Boltzmann constant.) 
The relationship \eqref{eq:canonical_energy} originates in statistical
mechanics from the context of a system with a bath.  However, it is widely used in the study of
thermalization of isolated (bath-less) quantum systems, to obtain an energy-temperature
correspondence \cite{Rigol_thermalization_Nat2008,Rigol_thermalization_PRL2009,
  Rigol_Quenches_PRA2009,RigolSantos_Chaos_PRA10,
  SantosRigol_Thermalization_PRE2010,Roux_quantumquenches_PRA2010,
  RigolSrednicki_Thermalization_2012,SantosPolkovnikovRigol_typicality_PRE2012,
  NeunhahnMarquardt_thermalization_PRE2012,RigolSrednicki_FDT_PRL2013,
  SorgVidmarHeidrichMeisner_thermalization_PRA14,NandkishoreHuse_MBL_AnnRev2015,
  FratusSrednicki_EstateThermalization_PRE2015,Essler_Quench_JoSM2016,
  Greinergroup_thermalization_Science2016,AlessioRigol_Chaos_AdvPhys2016,
  GarrisonGrover_SingleEstate_PRX2018,LuGrover_RenyiEntropy_PRE2019,
  SekiYunoki_thermal_PRR2020,Noh_ETH_PRE2021}.  

An alternate way to define temperature is to use the thermodynamic relation
\cite{Reif_StatPhys, Reichl_StatPhys, huang_statisticalmech, RKPathria_StatPhys, LandauLifshitz_StatPhys, 
	kardar_statphysp_2007}
\begin{equation}\label{eq:entropy_temp}
	T = \left( \pd{E}{S} \right)_{X_i} = \left( \pd{S}{E} \right)_{X_i}^{-1}.
\end{equation}
Here $S$ is the (thermal) entropy, and the subscript $X_i$ denotes the system parameters that should 
be held constant.  For an isolated (i.e. microcanonical) quantum system, defining the entropy $S(E)$
at a particular energy $E$ involves counting the number of eigenstates (`microstates') within an
energy window $\dE$ around that energy $E$.  This raises the question of how to choose the width
$\dE$, or possibly avoiding an explicit choice of $\dE$ and instead estimating the density of
states.  In the large-size (thermodynamic) limit, these choices can be shown to become inconsequential.
This paper aims to explore the consequences of these choices for finite sizes, focusing
on systems of Hilbert space dimensions $\sim10^4$ typical for full numerical diagonalization
studies.  

Motivated by the analysis of Ref.~\cite{Gurarie_Equivalence_AJoP2007}, we consider four 
choices for defining the entropy.  In each case, we compare the resulting temperature obtained using Eq.\
\eqref{eq:entropy_temp} with the canonical temperature obtained directly from inverting Eq.\
\eqref{eq:canonical_energy}.

\begin{figure}
  \includegraphics[width=0.98\linewidth]{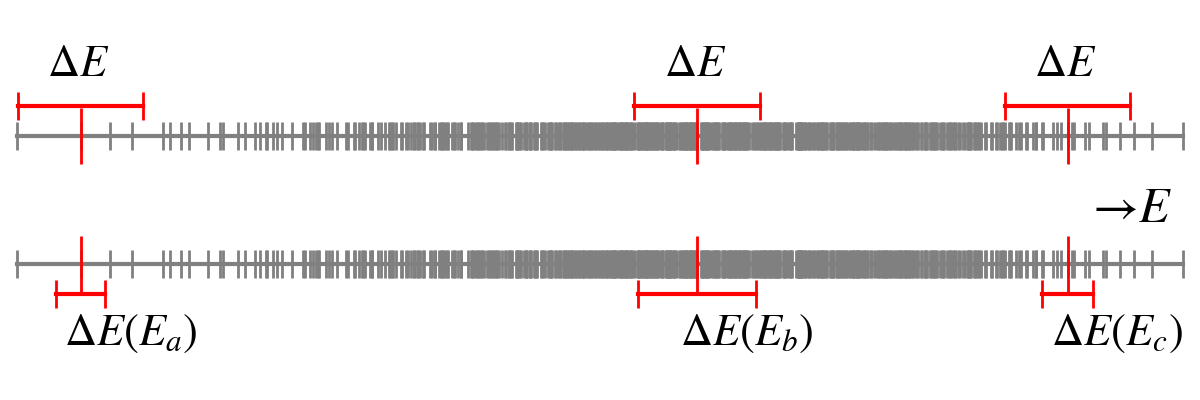}
  \caption{Illustration of two ways of choosing the energy window $\dE$ used to
    define the microcanonical entropy.  The top figure shows the obvious choice: the width $\dE$ is
    the same at all energies.  The lower schematic illustrates an energy-dependent width:
    $\dE(E)\propto\sqrt{T_c^2 C_c}$, where the temperature $T_c$ is the canonical temperature
    corresponding to energy $E$ and  $C_c$ is the heat capacity at $T=T_c$.  The window is shown at
    three energy values $E_a$, $E_b$, and $E_c$, which are in the regions of low temperature, infinite
    temperature, and low negative temperature.  Each vertical tick marks an eigenvalue. 
\label{fig:dE_illustration}
}
\end{figure}

First, we consider counting eigenstates in an arbitrarily chosen but constant (energy-independent) 
width window, as illustrated in Figure \ref{fig:dE_illustration}(top).  
This is the most obvious choice but turns out to be far from
optimal --- the resulting temperature deviates strongly at finite sizes from the canonical
temperature.  

Second, noting that the leading correction to the large-size limit can be made to disappear by
choosing $\dE\propto\sqrt{T_c^2C_c}$ \cite{Gurarie_Equivalence_AJoP2007}, we examine the result of
using such an energy-dependent window width.  (Here $C_c$ is the heat capacity.) Such an
energy-dependent window is illustrated in Figure \ref{fig:dE_illustration}(bottom). We show that this
choice works extremely well for the sizes of interest, modulo some caveats regarding the
proportionality constant.

Finally, we can compute the microcanonical entropy using standard numerical 
estimation procedures for the d.o.s.\ via approximations to the integrated d.o.s.\ or 
cumulative spectral function. We first formulate the definition without reference to a 
specific energy window $\dE$ and show that this choice is sub-optimal. We analyze the 
reason for the strong finite-size mismatch in this case. Secondly, we show that using 
the energy-dependent $\dE\propto\sqrt{T_c^2C_c}$, designed to account for finite-size 
effects, results in excellent agreement between the temperatures even at the small 
sizes investigated, without any fine-tuning of proportionality constants.
%

This article is organized as follows. Section \ref{sec:preliminaries} recounts the definitions and
saddle-point expressions that lead to the various choices for calculating the microcanonical entropy.
We also provide details (\ref{sec:numerics_Hamiltonians}) of the quantum many-body systems that we
use for numerical exploration --- we provide results for multiple systems with different geometries,
to ensure that the resulting conclusions are not artifacts of a particular lattice or Hamiltonian.
In the next three sections, we describe the procedures for, and the results of, the 
four ways of calculating entropy.  
First, Section \ref{sec:const_dE} describes counting eigenstates in a 
constant-width energy window $\dE$.  Second, Section \ref{sec:energy_dependent_dE} describes counting 
eigenstates in an energy-dependent window $\dE(E)$, designed to cancel out the explicit $\dE$-dependence 
of the resultant temperature, following/extending the suggestion of 
Ref.~\cite{Gurarie_Equivalence_AJoP2007}.  
Third, Section \ref{sec:Omega} outlines using a smoothened cumulative d.o.s.\ 
$\Omega(E)$ to calculate the d.o.s. $g(E)$. We can then choose either to neglect $\dE$, which we explain
leads to deviations, or make use of the derived energy dependent $\dE(E)$ designed to account for these deviations. 
Section \ref{sec:concl} provides concluding discussion and some context.


\section{Preliminaries}\label{sec:preliminaries}


We first recall the standard definition of microcanonical entropy and highlight the roles of the
density of states $g(E)$ and of the energy window $\dE$ (subsection \ref{sec:entropy_dos_window}).
We then recall (\ref{sec:saddlepoint}) the saddle-point formulation often used to show the
equivalence of microcanonical and canonical ensembles in the large-size limit
\cite{Gurarie_Equivalence_AJoP2007}, and extend beyond the leading order in order to analyze the
effect of $\dE$ at finite sizes.  Subsection \ref{sec:numerics_Hamiltonians} describes the quantum
systems to be used in subsequent sections to provide numerical examples.


\subsection{Entropy, density of states,  and the energy window} \label{sec:entropy_dos_window}

The microcanonical entropy $S(E)$ of a system at energy $E$ is  \cite{Gurarie_Equivalence_AJoP2007,
  Reichl_StatPhys, huang_statisticalmech, RKPathria_StatPhys, LandauLifshitz_StatPhys, kardar_statphysp_2007} 
\begin{equation}\label{eq:entropy_def}
	S(E) = k_B \ln\Gamma(E),
\end{equation}
where $\Gamma(E)$ is the statistical weight, which is the number
of microstates at energy $E$.  For a quantum system, microstates are to be interpreted as
eigenstates. For a quantum system with a discrete spectrum, counting the number of eigenstates is
problematic because at any particular energy there is usually zero or one eigenstate, or perhaps a
handful if there are degeneracies.  Thus, $S(E)$ would be $=-\infty$ for all values of energy except
at a countable number of discrete energy values.  This issue is
usually resolved \cite{Gurarie_Equivalence_AJoP2007, 
  RKPathria_StatPhys, LandauLifshitz_StatPhys} by taking
$\Gamma(E)$ to be the number of eigenstates in an energy window $\dE$ around $E$, rather than the
number of eigenstates exactly at energy $E$. Thus we define
\begin{align}
	\Gamma(E) &= \int_{E-\dE/2}^{E+\dE/2} g(E')dE' \notag \\
	&= \int_{E-\dE/2}^{E+\dE/2} \sum_{n}\delta(E'-E_n) dE', \label{eq:dos_int}
\end{align}
where the sum is understood to to include all the eigenvalues of the system Hamiltonian that lie in the window, 
i.e., all $E_n$ satisfying $E_n\in(E-\frac{\dE}{2},E+\frac{\dE}{2})$
\footnote{It is also common to use the window $[E,E+\dE)$. We choose to work with the window $[E-\dE/2,E+\dE/2]$. 
This choice presumably does not have significant effects.}.
It is then common to approximate this integral \cite{Gurarie_Equivalence_AJoP2007, Polkovnikov_DEntropy_AoP2011, SantosPolkovnikovRigol_Entropy_PRL2011, RussomannoFavaHeyl_Entropy_PRB2021, SilvaGoold_Temperature_PRA2022} via 
\begin{equation}\label{eq:dos}
	\Gamma(E) \approx \dE g(E) = \dE\sum_{n} \delta(E-E_n).
\end{equation}
%

Here, $g(E)$ is the density of states --- the number of many-body eigenstates per 
unit energy interval. Although defined as a sum over delta functions, the density of states can be 
thought of as a smooth function of energy over energy scales much larger than the typical level 
spacing. In numerical work, this is often achieved by broadening the delta functions into Gaussians or
Lorentzians of finite width \cite{Bruus_dAuriac_2DHubbard_levelstatistics_PRB1997,
  Molina_Relano_Retamosa_misleading_PRE2002, Duine_MacDonald_vortex_PRA2010,
  SorgVidmarHeidrichMeisner_thermalization_PRA14, Shchadilova_Ribeiro_Haque_PRL2014,
  Wehling_excitons_PRB2017, Pixley_etal_disorderedWeyl_PRB2017, LiuLu_GrapheneNanoribbons_PRR2022,
  HonerkampKennesNag_multiWeylSemimetals_PRB2022}. Alternatively, we can define $\Omega(E)$ as the
number of eigenstates with energy less than $E$, i.e., the integrated density of states.  Fitting a
smooth function to the staircase form of $\Omega(E)$, one can obtain a smooth density of states as
the derivative; $g(E)= \Omega'(E)$.

The entropy now depends on an energy window $\dE$, so we are thus faced with choosing an appropriate
$\dE$. The purpose of introducing a finite energy width was to smooth out the discreteness of the 
energy spectrum, so $\dE$ should be large enough to include a large number of eigenstates.  On the 
other hand, we want $\dE$ to be sufficiently small so that the density of states (regarded as a 
smooth function of energy) does not vary appreciably within the window, i.e., $\dE$ should be much 
smaller than the scale of the bandwidth of the system. Other than these general principles, we have 
the freedom to choose $\dE$, and in general, the entropy $S$ will depend on the choice. 
  
As we will explain next (\ref{sec:saddlepoint}), the sub-leading contributions to the entropy, which 
contain $\dE$, vanish in the large system size limit. So, for infinite systems, it will generally 
not matter what $\dE$ is, but the choice can drastically affect the entropy and resultant temperature 
for finite quantum systems. This paper aims to investigate and clarify the effect of this choice for 
finite systems whose Hilbert space sizes make them accessible to full numerical diagonalization.

\subsection{Saddle point expressions} \label{sec:saddlepoint}

To understand the role of $\dE$, it is helpful to express the entropy as an integral over (complex)
inverse temperature and perform saddle-point approximations \cite{Gurarie_Equivalence_AJoP2007,
  Reif_StatPhys, huang_statisticalmech}, extending the order beyond what is necessary in the
thermodynamic limit, to account for finite sizes.

Replacing the delta function in Eq.~\eqref{eq:dos} with
$\delta(x) = \int_{-\infty}^{\infty}(2\pi)^{-1}e^{i\beta x}d\beta$, and defining the free energy as
$F(\beta) = -\beta^{-1}\ln(\sum_n e^{-\beta E_n})$, we can write the entropy as
\begin{equation}\label{eq:entropy_free_def}
e^{S/k_B} = \Gamma(E) = \dE \int_{-i\infty}^{i\infty} \frac{d\beta}{2\pi} e^{\beta (E-F(\beta))} .
\end{equation}
To apply the saddle-point approximation, one first finds the critical point of the exponent  $h(\beta) =
\beta(E-F(\beta))$. The condition is 
\begin{equation}\label{eq:saddlepoint_condition}
E-F(\beta)-\beta\pd{F}{\beta}=0 \ ,
\end{equation}
which is equivalent to Eq.~\eqref{eq:canonical_energy} defining the canonical temperature.  Thus
the saddle point is at $\beta=\beta_c$, the canonical inverse temperature.  The leading order
saddle-point approximation is thus
\begin{equation}\label{eq:saddlepoint_leading}
\frac{S}{k_B} = \beta_c E- \beta_c F(\beta_c)
\end{equation}
with $\beta_c$ being the solution of Eq.~\eqref{eq:saddlepoint_condition} or Eq.~\eqref{eq:canonical_energy}. 
This matches the standard thermodynamic relation, and is consistent with the energy 
derivative of $S$ being the inverse temperature $\beta_c$.  

To examine the effect of $\dE$, one needs to extend the calculation to the next order.  One expands
$h(\beta)$ as a Taylor series about $\beta_c$, up to second order, as the first order term is zero
by definition.  Introducing the heat capacity $C = \pd{E}{T} = -T\pds{F}{T}$,
%
%
one obtains \cite{Gurarie_Equivalence_AJoP2007}
\begin{equation}\label{eq:entropy_sub}
e^{S/k_B} \approx \frac{\dE}{2\pi} e^{\beta_c(E-F(\beta_c))}\int_{-\infty}^{\infty} dy e^{-k_B
  T_{c}^{2}C_c y^2/2}.
\end{equation}
Here, $T_{c}$ and $C_{c}$ are the values at the saddle point $\beta_c$.  Evaluating the  Gaussian
integral, one obtains 
\begin{multline} \label{eq:entropy_final}
\frac{S}{k_B} =  \beta_c E- \beta_c F(\beta_c)  + \ln\dE -  \ln\sqrt{2\pi k_{B}T_{c}^{2}C_{c}} . 
\end{multline}
Here $\beta_c$ is determined by the energy, and hence so are $T_{c}$ and $C_{c}$. The two leading
terms are both extensive in system size. The first correction term $\ln\dE$ is sub-extensive as long
as $\dE$ is not chosen to grow exponentially or faster with system size. The second correction term
grows logarithmically with system size, as the heat capacity is extensive.
However, as it appears in the argument of a logarithm, its dependence on $L$ 
disappears when differentiated with respect to $E$. This means that when differentiating 
Eq.~\eqref{eq:entropy_final} to obtain $\beta$, the contribution from the last term does 
not grow with $L$, not even logarithmically.
Thus, at large enough sizes, the leading-order saddle-point approximation suffices, 
and the choice of energy window $\dE$ plays no role. 

However, the two correction terms should be considered when calculating the entropy at 
finite sizes. From Eq.~\eqref{eq:entropy_final}, we see that choosing $\dE$ to be an energy-independent 
constant (the most obvious choice, explored in Section \ref{sec:const_dE}) would leave an 
energy-dependent correction term, causing deviations from the canonical temperature. We also see (as 
pointed out in Ref.~\cite{Gurarie_Equivalence_AJoP2007} and explored below in Section 
\ref{sec:energy_dependent_dE}) that the energy dependence of the correction terms could be canceled by a 
judicious choice of energy dependence for the width $\dE$.  


\subsection{Hamiltonians and numerics}  \label{sec:numerics_Hamiltonians}

To investigate the effect of different choices for defining the microcanonical entropy in finite systems, 
we use several spin-1/2 lattice systems consisting of $L$ spins, $N$ of which are up. The spins interact 
via \textit{XXZ} interactions, which have a $U(1)$ symmetry conserving $N$. We check all results and show 
data for 1D, 2D, and fully-connected geometries, to demonstrate that our results are very general and not 
particular to any model. In the case of the 1D chain, we also include magnetic fields in the $z$ and $x$ 
directions; the latter break the $U(1)$ symmetry. The Hilbert space dimension is $D=\binom{L}{N}$ when $N$ 
is conserved and $D=2^L$ otherwise. 

The system parameters are always chosen such that the level spacing statistics match that expected of chaotic 
quantum systems. This ensures there are no complications due to proximity to integrability or localization.

\textit{Staggered field XXZ chain}: Starting with the open-boundary \textit{XXZ} chain (anisotropic 
Heisenberg chain)
\begin{align} \label{eq:Ham_chain}
	H_H = J \sum_{j=1}^{L-1} (S^{x}_{j} S^{x}_{j+1} + S^{y}_{j} S^{y}_{j+1}) + 
	\Delta \sum_{j=1}^{L-1} S^{z}_{j} S^{z}_{j+1}
\end{align} 
we introduce magnetic fields in both the transverse ($x$) and longitudinal ($z$) directions. To remove 
symmetries in the model we stagger the two fields along the even and odd sites respectively, and in addition, 
we break the staggered pattern at the start of the chain by inserting $x$ and $z$ fields on the first and 
second sites respectively:
\begin{align}\label{eq:Ham_stag}
	H_S = H_H + \sum_{j} h_z S^{z}_{2j+1} + \sum_{j} h_x S^{x}_{2j} +h_{x}S^{x}_{1} + h_{z}S^{z}_{2}. 
\end{align}

\textit{Square Lattice}: This model is a two-dimensional square lattice with open boundary
conditions and \textit{XXZ} interactions:
\begin{align}  \label{eq:Ham_square}
	H_{\text{SQ}} = \sum_{\langle j,k \rangle} J_{jk}(S^{x}_{j} S^{x}_{k} + S^{y}_{j} S^{y}_{k}) +  
	\Delta_{jk} S^{z}_{j} S^{z}_{k}
\end{align} 
where $\langle\rangle$ means that we restrict the summation to nearest-neighbor pairs.  In order to
remove any symmetries, we draw the values $J_{jk}$, $\Delta_{jk}$ from the uniform distributions
$[0,2]$, $[0,1]$ respectively.

\textit{Fully connected Lattice}: This model consists of a network of spins, where each spin $j$ is
connected to every other spin $k$:
\begin{align}  \label{eq:Ham_FC}
	H_{\text{FC}} = \sum_{j}  \sum_{k\neq j} J_{jk}(S^{x}_{j} S^{x}_{k} + S^{y}_{j} S^{y}_{k}) +  
	\Delta_{jk} S^{z}_{j} S^{z}_{k}
\end{align}  
We draw the values $J_{jk}$, $\Delta_{jk}$ from the uniform distribution $[-0.4,0.4]$ and
$[-0.1,0.1]$ respectively.

\textit{Units}: For numerical work, we use $k_B=1$. We express energies in units of spin
couplings. Thus, if the coupling $J$ (or $J_{jk}$) is set to 1, when we plot energy values, 
they are $E/J$; whereas if the couplings have other values, then the energy is to be thought of as the 
energy divided by a hypothetical coupling equal to $1$.  In this way, the units of all energies, 
temperatures, and entropies presented in the figures are determined.

\section{Constant width window}\label{sec:const_dE}

\begin{figure}
	\includegraphics[width=0.98\linewidth]{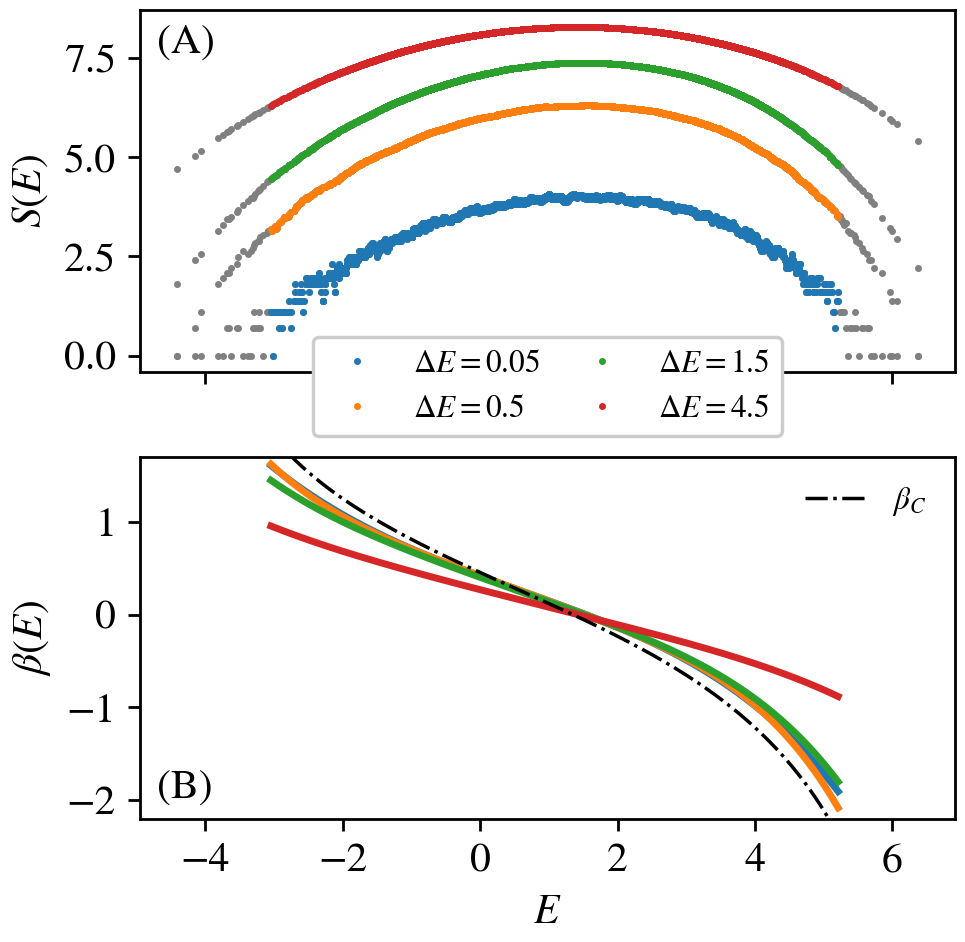}
	\includegraphics[width=0.96\linewidth]{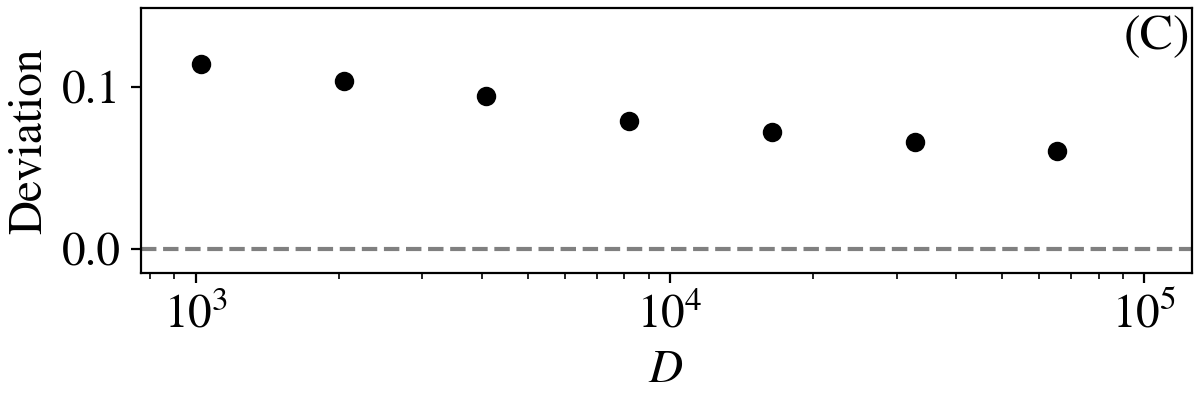}
	\caption{\textbf{(A)} Microcanonical entropy $S(E)$ calculated with constant (energy-independent)
	  $\dE$.  Evaluated only at the eigenenergies $E=E_n$.  \textbf{(B)} Resultant inverse temperature
	  $\beta(E)$ obtained as derivative of 6th-order polynomial fitted to $S(E)$ data.  25 eigenstates
	  from both ends of the spectrum (shown in gray in \textbf{(A)}) were excluded from the fit.  \textbf{(A,B)}
	  $5\times$4 square \textit{XXZ} lattice, Eq.\ \eqref{eq:Ham_square}, with $L=5\times4$,
	  $N=4$. \textbf{(C):} Root-mean-square (RMS) distance between $\beta(E)$ and the canonical
	  temperature, versus Hilbert space dimension $D=2^L$, with $\dE = 0.5$.  Spin chain with $L$ spins,
	  Eq.\ \eqref{eq:Ham_stag}, $J=1$, $\Delta=0.95$, $h_z = h_x = 0.5$.  Units used in this and all
	  subsequent figures are explained in subsection \ref{sec:numerics_Hamiltonians}.  }
		\label{fig:S_const_dE}
\end{figure}

We start by discussing the obvious choice for defining entropy --- counting eigenstates in a constant
energy-independent width $\dE$. Eq.\ \eqref{eq:entropy_final} implies that an energy-independent
$\dE$ acts as an energy-independent shift of the entropy, and hence does not affect the temperature.
However, the last term in Eq.\ \eqref{eq:entropy_final} is energy-dependent, hence the temperature
will differ from the canonical temperature which corresponds to the leading (first two) terms. This
deviation should vanish in the large-size limit, but the extent of this deviation for finite sizes
is not a priori clear.
We show below using explicit numerical examples that using an energy-independent width is a poor
choice for the system sizes under study here (sizes accessible to exact diagonalization), as the
resulting temperature deviates significantly from the canonical temperature.

Examples of entropy found by counting the number of eigenvalues within the window
$[E-\dE/2,E+\dE/2]$, are shown in Figure \ref{fig:S_const_dE}(A) for an open-boundary square
spin-$\frac{1}{2}$ lattice with random \textit{XXZ} interactions. Here and in later figures, we
plot one point for each eigenvalue, so that there is at least one eigenvalue in each window --- the
minimum value of entropy is $\ln1=0$ and we thus avoid the possibility of obtaining entropy
$\ln0=-\infty$.  

To calculate the temperature from the entropy, we fit a polynomial to the $S(E)$ points and then
take the derivative of the polynomial; results are shown in  Figure
\ref{fig:S_const_dE}(B).

Increasing $\dE$ by a factor approximately increases the number of eigenvalues in each window by
that factor, so that the $S(E)$ curve undergoes an approximately constant upward shift.  This leaves
the temperature $[S'(E)]^{-1}$ approximately unchanged.  Accordingly, for moderate $\dE$, the
calculated $\beta(E)$ curves are robust to changes in $\dE$.  For the largest $\dE (=4.5)$ shown,
this argument does not work as the window is significant compared to the variation scale of the
density of states $g(E)$, so that $g(E)$ cannot be considered constant within
each window.  This leads to a markedly different $\beta(E)$ curve for the very large $\dE$ case.  This effect is not captured by the saddle-point analysis or Eq.\ \eqref{eq:entropy_final}.  

Even for moderate $\dE$, the deviation from the canonical $\beta_C(E)$ curve is considerable. This
shows that using a constant energy window width is not a feasible approach to defining entropy for
finite quantum systems having Hilbert space dimensions $\sim{}O(10^4)$. The results of Figure
\ref{fig:S_const_dE}(B) provide a visual presentation of the extent of the discrepancy for such sizes. In Figure
\ref{fig:S_const_dE}(C) we show the root-mean-square deviation of the inverse temperatures obtained
with a constant $\dE$ from the canonical $\beta_C(E)$ values, as a function of Hilbert space
dimension. (We show data for a spin chain, for which finite size scaling is numerically more
convenient than for a square lattice.) The deviation decreases with increasing size, as expected, but
very slowly.

A remark on the edges of the spectrum is in order.  In the edge regions, as there are only a few
eigenvalues in each window, the values of entropy are noticeably discrete, $\ln n$ where $n$ is a
small integer.  For a regime with such discrete behavior, statistical-mechanical considerations are
not meaningful --- we expect entropy to be a continuous function of energy.  We therefore omit the
edges from the polynomial fit to $S(E)$.  In Figure \ref{fig:S_const_dE} we have omitted 25
eigenstates from each edge of the spectrum.  This is approximately the energy region for which
$S(E)\lesssim\ln10$ for the $\dE=0.5$ case.  Later in the paper, the same criterion will be used to
exclude the spectral edges.

\section{Energy dependent window}\label{sec:energy_dependent_dE}

We now describe counting eigenstates in an energy-dependent $\dE\propto\sqrt{T_c^2C_c}$.

\subsection{Rationale}

In Eq.\ \eqref{eq:entropy_final}, the leading-order terms (first two terms) yield the canonical
temperature upon differentiation by energy, but there are two correction terms, $\ln\dE$ and
$-\ln\sqrt{2\pi{k_B}T_cC_c}$.  Ref.\ \cite{Gurarie_Equivalence_AJoP2007} suggests choosing the two
terms to be equal so that the corrections vanish and the canonical temperature is recovered.

We note, however, that the correction terms need not cancel exactly --- they will not affect the
temperature $[S'(E)]^{-1}$ as long as they sum to an energy-dependent constant.  Thus we could
choose
\begin{equation}\label{eq:eps_def}
\dE(E) \equiv \alpha^{-1}\sqrt{2\pi k_{B}T_{c}^{2}C_{c}}.
\end{equation} 
with $\alpha$ being some constant.  The window $\dE$ is then energy-dependent because $T_c$ and
$C_c$ are energy-dependent.  The proposal of Ref.\ \cite{Gurarie_Equivalence_AJoP2007} corresponds
to $\alpha=1$.  

The energy-dependent $\dE$ can also be motivated by the physical idea that the energy window should be determined by the fluctuation of energy in the canonical ensemble, as suggested, e.g., in Refs.~\cite{Polkovnikov_DEntropy_AoP2011,SantosPolkovnikovRigol_Entropy_PRL2011}.  It turns out \cite{huang_statisticalmech} that the distribution in energy in the canonical ensemble has width $\propto\sqrt{T^2C}$.  Thus, choosing the energy uncertainty as the energy window leads to the same prescription as Eq.\ \eqref{eq:eps_def}.

Labeling the number of states in an energy-dependent energy window $\dE(E)$ as $\tilde{\Gamma}(E)$,
the entropy is obtained as 
\begin{equation}\label{eq:S_epsilon}
\tilde{S}(E) = k_B \ln \left( \tilde{\Gamma}(E) \right) .
\end{equation}
Differentiation gives the corresponding inverse temperature $\tilde{\beta}$.   In the following
subsection we will describe doing this numerically, and discuss the results obtained.

We do not know of a principle guiding the choice of $\alpha$.  
We will show that, for the sizes of primary interest to us, $\alpha$ needs to be larger than $1$.  

The reason is that for $\alpha=1$, the energy window $\dE(E)$ turns out to be too broad, 
exceeding the energy scale at which the density of states varies, leading to poor results 
similar to the large constant $\dE$ case in Figure \ref{fig:S_const_dE}. 
This is clear from Eq.~\eqref{eq:eps_def}; the right-hand side scales like $\sqrt{L}$ 
as $C_c$ is extensive, while the variance of the density of states typically scales linearly with 
$L$. Thus, for the system sizes under consideration here, $\sqrt{L}$ can not be considered negligible 
compared with $L$.
For larger system sizes, the acceptable range of $\alpha$ broadens,
compatible with our expectation that the choice of $\dE$ should be less important for larger sizes.

\subsection{Numerical calculation and results}\label{sec:fin_sys_res}

\begin{figure}
  \includegraphics[width=0.95\linewidth]{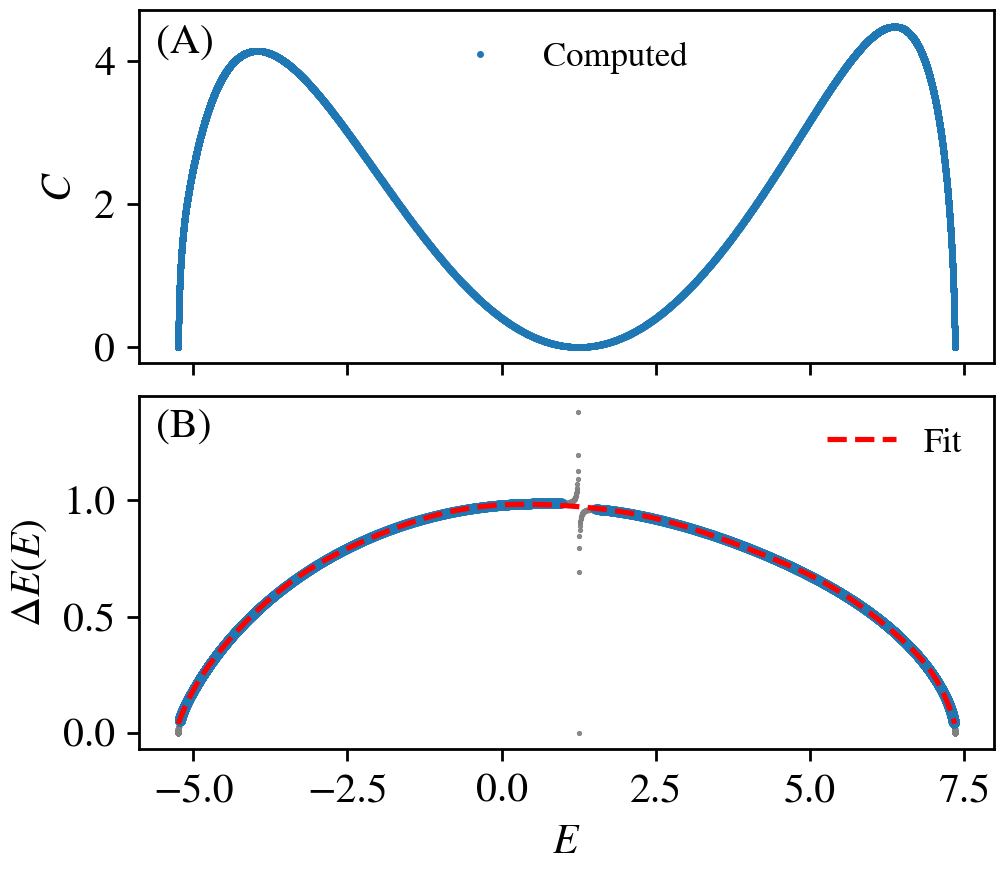}
\caption{\textbf{(A):} Specific heat $C = \partial E/ \partial T$.   \textbf{(B):} $\dE(E) =
  \alpha^{-1}\sqrt{2\pi k_B T_{c}^{2}C_{c}}$, for $\alpha=5$, numerically computed.  Dashed curve: 12-th order
  polynomial fit, used to avoid the spurious divergence.  Data for  $5\times$4 square \textit{XXZ}
  lattice, Eq.\ \eqref{eq:Ham_square}, with  $N=4$.} 
\label{fig:de_fit}
\end{figure}

In order to obtain $\dE(E)$ as defined in Eq.\ \eqref{eq:eps_def}, we first need to calculate the heat capacity $C_c$ and the canonical temperature.  We start by numerically computing the energy $E$ as a function of $\beta$ (or of $T$) via Eq.~\eqref{eq:canonical_energy}, using the numerically calculated eigenvalues of the Hamiltonian $H$.  This leaves us with a (numerical approximation to a) smooth injective function $E(T)$.  As usual, inverting this numerically provides the canonical temperature $T_c$ as a function of $E$.  In addition, estimating the derivative of $E(T)$ provides us with the heat capacity $C = \partial E/ \partial T$, which we evaluate at the canonical value $T_c$.  Figure \ref{fig:de_fit}(A) shows an example of $C_c$ as a function of energy, calculated in this way.  The function is non-monotonic, going to zero at both zero and infinite temperatures.


We can now calculate $\dE(E)$ via \eqref{eq:eps_def} for all energies.  Figure \ref{fig:de_fit}(B) shows an example. 

At the center of the spectrum, $T_c\to\infty$ and $C_c\to0$, leading to a finite $\dE(E)\propto\sqrt{T_c^2C_c}$.  Due to the numerical discreteness of computed $E$ and $T$ functions, the computed $\dE(E)$ acquires a spurious divergence at this point.  This effect can be confined to a smaller energy region by using a finer grid of $T$ (or $E$) values.  We avoid the effect by fitting a polynomial, excluding points within the direct vicinity of the discontinuity.  An example  is shown in  Figure \ref{fig:de_fit}(B).  The fitted polynomial is then used as $\dE(E)$.

\begin{figure}
\hspace{0.1\linewidth}
\includegraphics[width=0.3\linewidth]{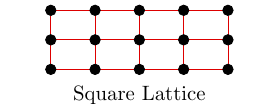}
\hspace{0.15\linewidth}
\includegraphics[width=0.3\linewidth]{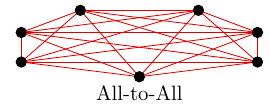}\\
\includegraphics[width=0.999\linewidth]{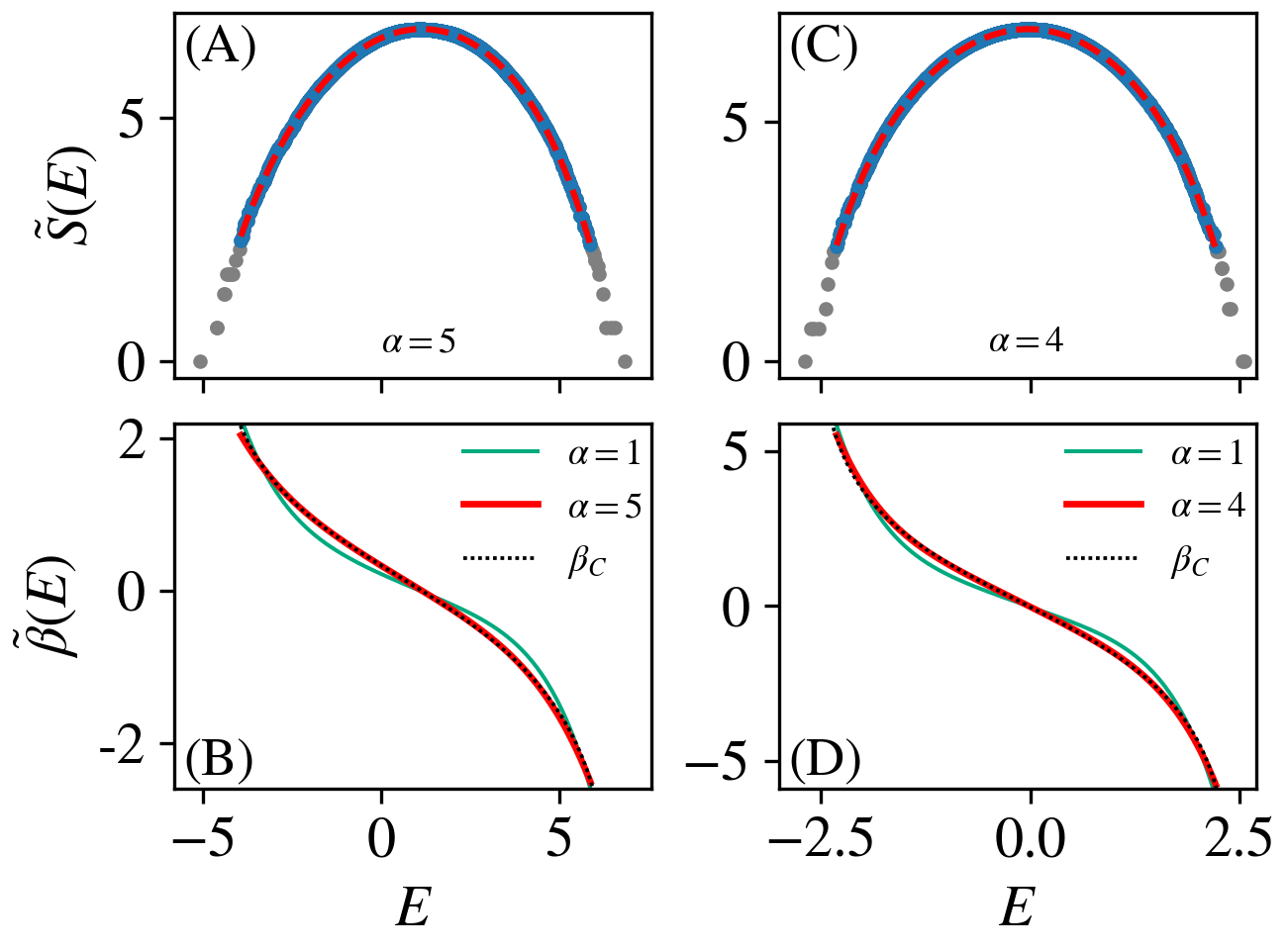}
\includegraphics[width=0.999\linewidth]{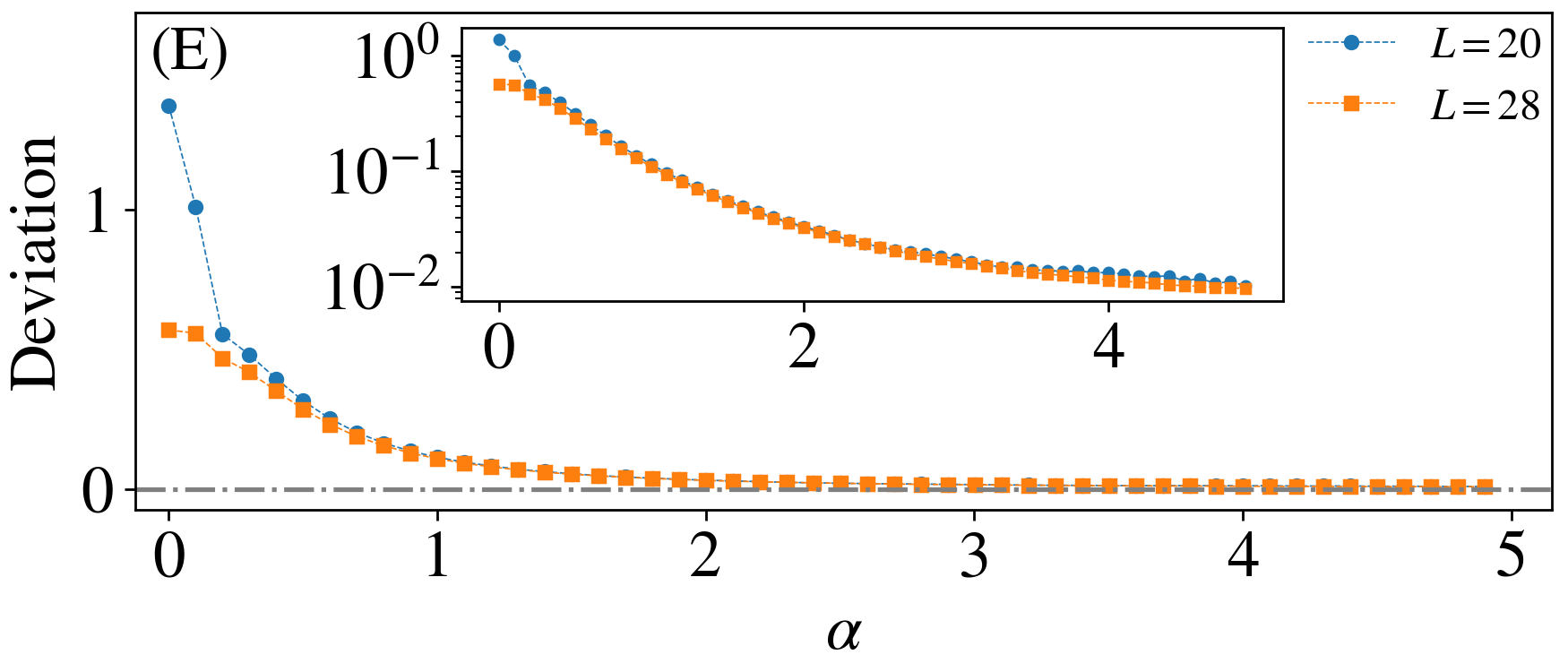}
\caption{Using an energy-dependent window, Eq.\ \eqref{eq:eps_def}.  Left panels \textbf{(A,B)} and 
  right panels \textbf{(C,D)} show data for two different systems.  \textbf{(A,C):} Microcanonical
  entropy calculated with $\dE(E) = \alpha^{-1}\sqrt{2\pi k_B T_{c}^{2}C_{c}}$ along with 6th-order
  polynomial fit.  Fit excludes gray points.  \textbf{(B,D)} Resultant temperature
  $\tilde{\beta}(E)$, with two different values of $\alpha$ in each case, compared with canonical
  temperature $\beta_C(E)$.  \textbf{(A,B)} Data for $5\times$4 square \textit{XXZ} lattice, Eq.\
  \eqref{eq:Ham_square}, with $N=4$.  \textbf{(C,D)} Fully connected lattice, , Eq.\
  \eqref{eq:Ham_FC}, with $L=16$, $N=5$.  \textbf{(E)} RMS distance between $\tilde{\beta}(E)$ and
  $\beta_C(E)$ plotted against $\alpha$ for two square lattices, Eq.\ \eqref{eq:Ham_square}, a
  $5\times4$ lattice with $N=4$, and a $7\times4$ lattice with $N=5$.  Inset: logarithmic scale.
}\label{fig:S_eps}
\end{figure}

We are now equipped to compute the entropy using an energy dependent energy window, $\dE(E) = \alpha^{-1}\sqrt{2\pi k_B T_{c}^{2}C_{c}}$.
In Figure \ref{fig:S_eps}(A,C), we show the numerically computed entropy $\tilde{S}$. The data shown is for a chaotic square lattice of spins-1/2's with \textit{XXZ}-like connections between nearest neighbor spins, and also for a fully connected lattice with \textit{XXZ}-like connections between every pair of sites.  We used $\alpha=5$ and $\alpha=4$ in the two cases.   

As before, we numerically fit a polynomial to our entropy data, excluding values at the edge of the spectrum, and then take its derivative to obtain a temperature. The resultant inverse temperature $\tilde{\beta}$ are shown in Figures \ref{fig:S_eps} (B,D).  For  $\alpha\sim5$, the temperature matches the canonical temperature remarkably well. 

We found that the procedure provides an excellent match between $\tilde{\beta}$ and $\beta_c$ when $\alpha$ is larger than some minimum value, as shown by the vanishing root-mean-square deviation in Figure \ref{fig:S_eps}(E).  Beyond this minimal $\alpha$ the exact choice is somewhat arbitrary, as long as the window is not made ultra-small.  Figure \ref{fig:S_eps}(E) shows that this minimal value (the value of $\alpha$ at which the deviation drops to zero) is smaller for larger systems.   This is consistent with the expectation that the exact choice of the window $\dE$ is increasingly irrelevant as the system size is increased.  

To summarize: this numerical analysis shows that using an energy-dependent window is a very successful strategy for defining entropy in a finite-size system.  We observed good agreement
between the resultant temperature and the canonical temperature for all systems we checked,
including various 1D chaotic spin chains, and a handful of 2D lattices, including that presented in
Figure \ref{fig:S_eps}.



\section{Using the integrated density of states}\label{sec:Omega}
Here, we formulate the entropy in terms of the integrated d.o.s.\ $\Omega(E)$ and describe how to 
numerically obtain a smoothened approximation to the d.o.s.\ $g(E) = \Omega'(E)$. First, we approximate
the entropy by neglecting the $\dE$ contribution, and explain that this leads to finite-size deviations.
Second, we use the smoothened $\Omega'(E)$ with the energy-dependent $\dE(E)$ designed to account for 
the finite size deviations, and show that it leads to excellent agreement with the canonical temperature.

\subsection{Formulation}

An alternative to counting eigenstates in an explicit energy window is to use the expression for 
the entropy  in terms of the (integrated) density of states:
\begin{align}\label{eq:entropy_from_dos}
	S &= k_B \ln\Gamma(E) = k_B \ln\bigg(g(E)\dE\bigg) \notag\\ 
	&= k_B \ln g(E) + k_B \ln\dE \notag \\ 
	&=   k_B \ln(\pd{\Omega(E)}{E}) + k_B \ln\dE  .
\end{align}
Here,
\begin{equation}
  \Omega(E) = \sum_{n=1}^{D} \Theta(E-E_n)
\end{equation}
is the number of eigenstates $\ket{E_n}$ with energy less than $E$, i.e., the 
integrated d.o.s. or cumulative d.o.s., also known as the cumulative level density
\cite{Graef_etal_stadium_PRL1992, Relano_Dukelsky_Retamosa_integrable_PRE2004, 
	Le_etal_heliumatoms_PRA2005,
  	Santhanam_Bandyopadhyay_PRL2005, Mulhall_PRC2009, Wang_Gong_butterflyFloquet_PRE2010,
  	Enciso_etal_HaldaneShastry_PRE2010, Jain_Samajdar_billiards_RMP2017,
  	Ormand_Brown_nuclear_Lanczos_PRC2020, Corps_Molina_Relano_Thoulessenergy_MBL_PRB2020,
  	Magner_etal_semiclassical_PRC2021, Corps_Relano_longrange_PRE2021} 
or the cumulative spectral function 
\cite{Ciliberti_Grigera_localization_levelspacing_PRE2004, Gusso_Rego_nanoelectromechanical_PRB2006,
  SantosRigol_Thermalization_PRE2010, Frye_Hutson_approachchaos_PRA2016,
  Dettmann_Knight_randomgraphs_EPL2017, Abuelenin_unfolding_PhysicaA2018,
  Chaudhuri_etal_tensormodels_PRD2018, Vidmar_HeidrichMeisner_ETHpolaron_PRB2019,
  vonSmekal_2color_staggeredquarks_PRD2019, Corps_Molina_Relano_Thoulessenergy_MBL_PRB2020,
  Kourehpaz_TypicalityChaos_ent2022, Lydzba_Sowinski_fewfermions_PRA2022}.  

\begin{figure}
	\includegraphics[width=0.98\linewidth]{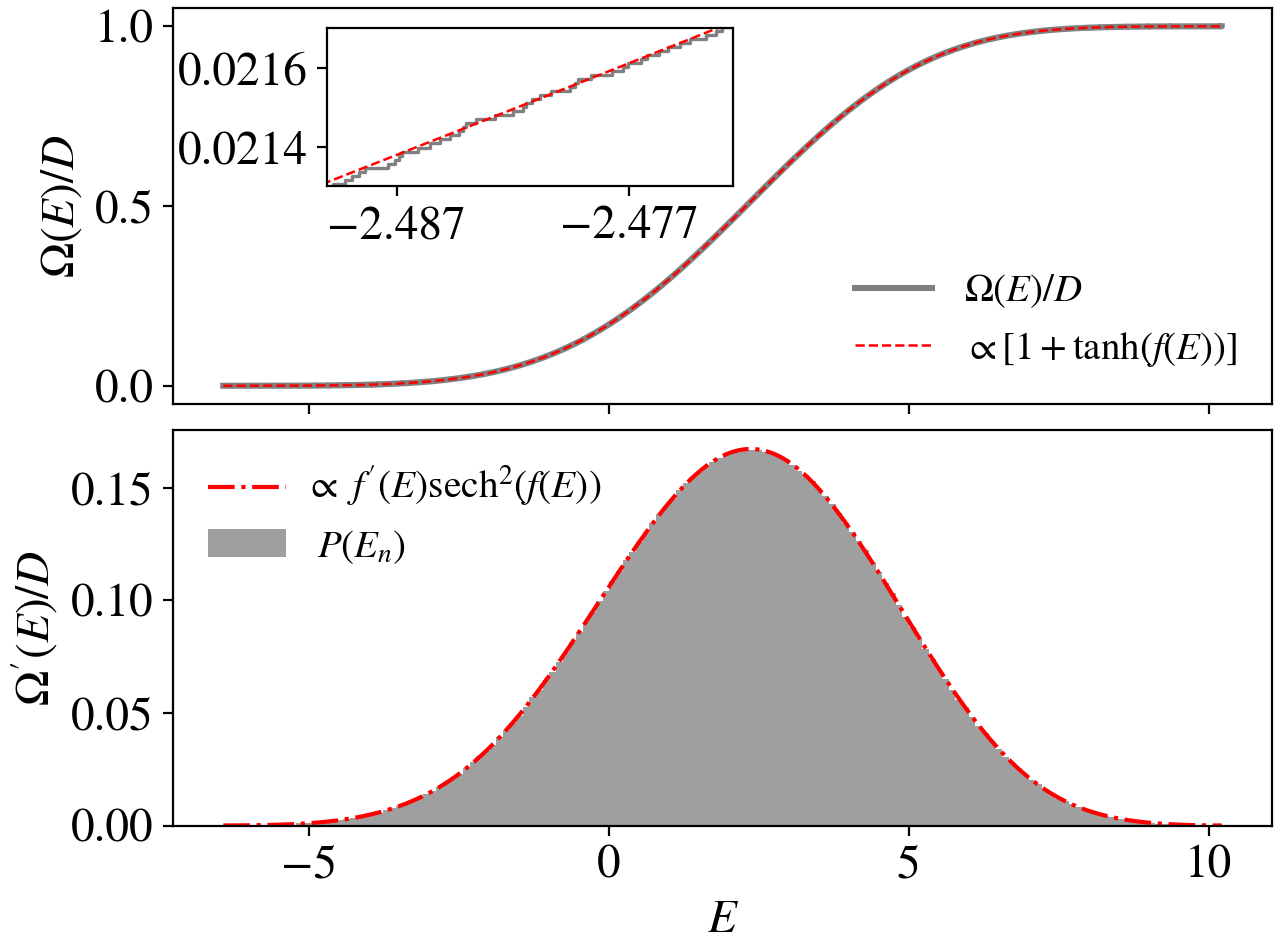}
	\caption{\label{fig:omega_fit}
	\textbf{Upper:} Cumulative density of states $\Omega(E)$ (normalized by Hilbert space
	  dimension),   fitted by a  polynomial  $f(E)$ 
	  of degree 10. Inset: zoom.  \textbf{Lower:}
	  $\Omega'(E)/D$ from derivative of fitted function, compared with 
	  histogram (distribution) of eigenvalues, 
	  $P(E_n)$.  Data for $7\times4$ square \textit{XXZ} lattice, 
	  Eq.\ \eqref{eq:Ham_square}, with $N=5$.  
	}
\end{figure}

From the eigenvalue spectrum of the system Hamiltonian, computed numerically using exact
diagonalization, the integrated d.o.s.\ $\Omega(E)$ can be obtained as a function of $E$. 
It is a series of steps with constant integer values between eigenvalues and a step 
to the next integer at every eigenvalue. 
In order to obtain a derivative of this non-smooth function, we will fit an 
analytic function to the computed
$\Omega(E)$ data, and then simply take its derivative.  Fitting an analytic function 
to $\Omega(E)$ is common practice in the unfolding procedure utilized for computing 
level spacing statistics
\cite{Haake1991QuantumSO,Guhr_RMTinQM_PR1998, Molina_Relano_Retamosa_misleading_PRE2002,
  Abul_Unfolding_PhysicaA2014, Abuelenin_unfolding_PhysicaA2018, Vidmar_HeidrichMeisner_ETHpolaron_PRB2019}.

We found that using a function of the form 
\begin{equation}
\frac{D}{2} \big[ 1+\tanh(f(E)) \big]
\end{equation}
works remarkably well to fit the numerical $\Omega(E)$ data, where $f(E)$ is 
some polynomial in $E$.
If $f(E)$ is a monotonically increasing polynomial, then the form of the 
function automatically imposes the correct low-energy and high-energy behavior 
of the smoothened $\Omega(E)$.  
This functional form was inspired by its use in \cite{Chaudhuri_etal_tensormodels_PRD2018} 
to unfold a many-body spectrum.  Once the fitting function $f(E)$ is determined, 
the density of states is obtained as the derivative: 
\begin{equation}\label{eq:dos_fit}
	g(E) = \Omega'(E) = \frac{D}{2}\sech^2(f(E))f'(E)
\end{equation}

An example of numerically calculated $\Omega(E)$ and the fitted function are shown in Figure
\ref{fig:omega_fit} (top).  
The resulting derivative $\Omega'(E)$ is compared with the normalized histogram of 
eigenvalues (bottom). In both panels, we normalize the functions by $D$, so that the function
itself is plotted between $0$ and $1$, and its derivative can be directly compared with the
normalized eigenvalue distribution $P(E_n)$. 
With $\Omega'(E)$ in hand, we are now equipped to compute the entropy and resultant temperature. 
In the following subsections we present two ways to do so.

\subsection{Using the integrated D.O.S. while avoiding $\dE$}\label{sec:omega_res}


The density of states increases exponentially with the system size; 
hence the first term in Eq.~\eqref{eq:entropy_from_dos} is extensive. 
On the other hand, $\ln\dE$ is presumably either constant or at most weakly 
increasing with system size. Thus, at sufficiently large system sizes the second 
term can be neglected so that the first term approximates the entropy:
\begin{equation}\label{eq:def-OmegaEntropy}
	S \approx S_\Omega  = k_B \ln g(E) = k_B \ln(\pd{\Omega(E)}{E}) .
\end{equation}
One can thus use a continuous approximation for $\Omega(E)$ to define a continuous $S_\Omega(E)$, without choosing an explicit energy window $\dE$.  


$S_\Omega$ is  the logarithm of $g(E) = \Omega'(E)$, and $\beta_{\Omega}$ is thus the 
derivative of this. For the analytical approximation to $\Omega'(E)$, Eq.~\eqref{eq:dos_fit}, 
one obtains
\begin{equation}
\beta_{\Omega}(E) = \frac{f''(E)}{f'(E)} - 2f'(E)\cdot\tanh(f(E)).
\end{equation}  

\begin{figure}
	\hspace{0.12\linewidth}
	\includegraphics[width=0.3\linewidth]{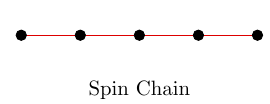}
	\hspace{0.15\linewidth}
	\includegraphics[width=0.3\linewidth]{Square_tikzit.pdf}\\
	\includegraphics[width=0.99\linewidth]{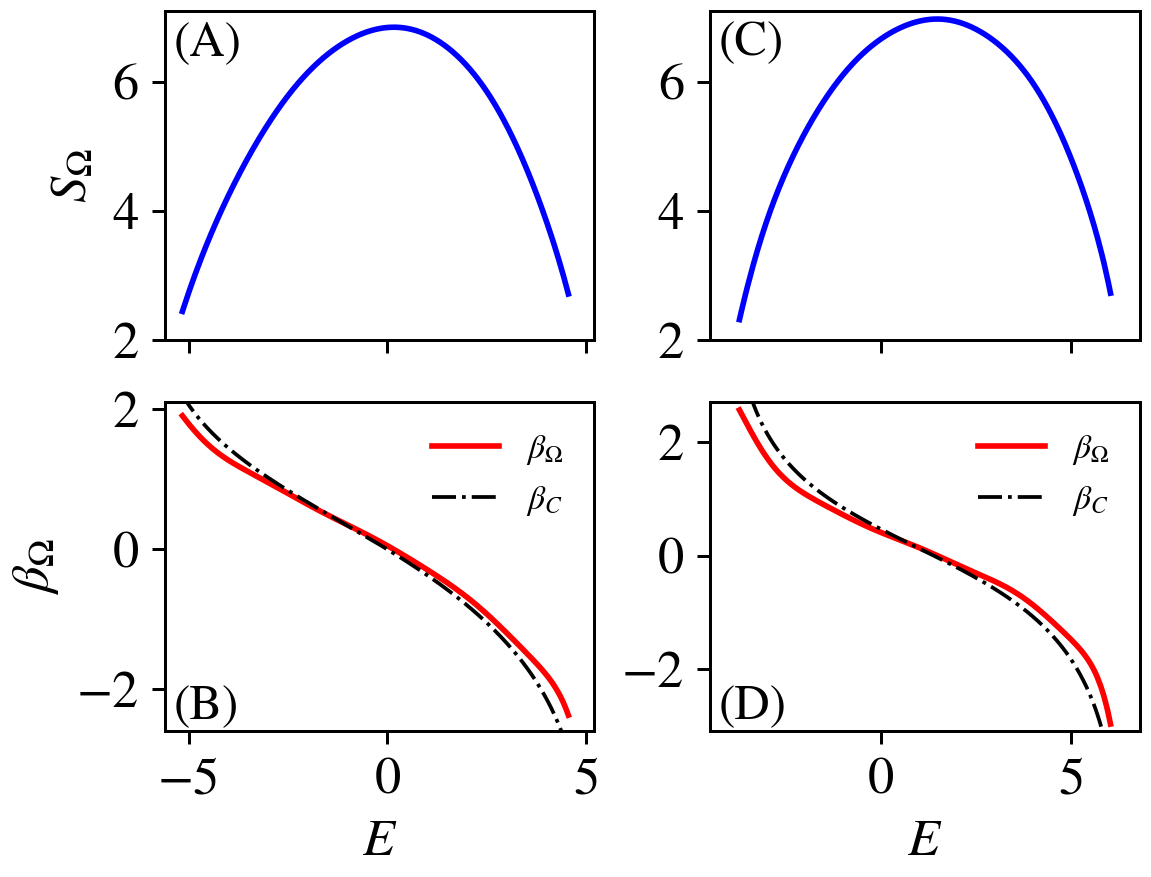}
	\includegraphics[width=0.97\linewidth]{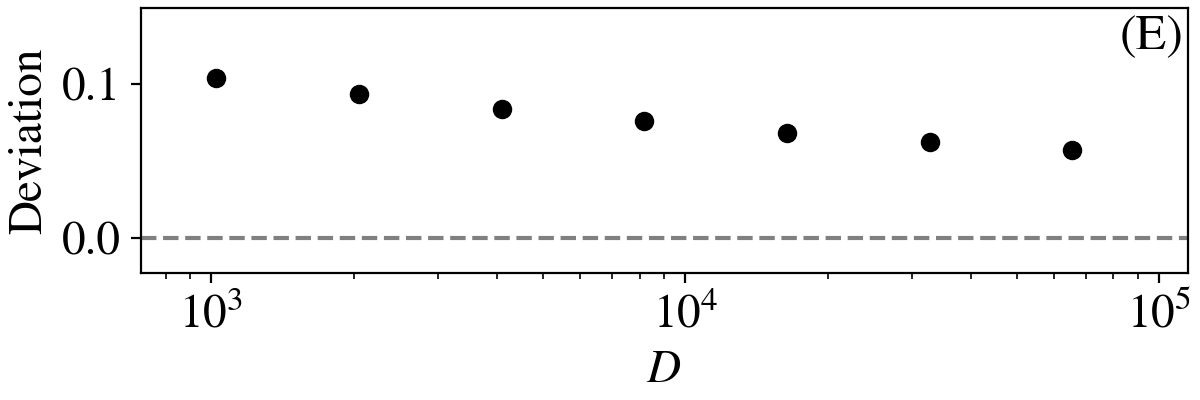}
	\caption{  \label{fig:S_omega}
	Entropy and temperature using a continuous approximation to the cumulative density of states. 
	\textbf{(A,C)} $S_{\Omega} = \ln\Omega'(E)$. 
	\textbf{(B,D)} Resultant inverse temperature $\beta_{\Omega} = \partial_E S_{\Omega}$ compared with
	the canonical temperature $\beta_C$. 
	\textbf{(A,B)} Staggered field \textit{XXZ} chain, Eq.\ \eqref{eq:Ham_stag},  with $L=12$, $J=1$, $\Delta=0.95$,
	$h_z = h_x = 0.5$.  
	\textbf{(C,D)} $5\times4$ square \textit{XXZ} lattice, Eq.\ \eqref{eq:Ham_square},  with $N=4$.
	\textbf{(E)} RMS distance between $\beta_{\Omega}(E)$ and $\beta_C(E)$, versus $D$ $(2^L)$, for the
	spin chain, Eq.\ \eqref{eq:Ham_stag}, with the same parameters as in \textbf{(A,B)} and variable $L$.
		}
\end{figure}

Examples of the computed entropy $S_{\Omega}$ are shown in Figure \ref{fig:S_omega}(A,C). 
The numerical results shown are for a 1D spin chain, and a square lattice, both 
with nearest neighbor \textit{XXZ}-like connections and open boundary conditions. 
The resultant inverse temperatures $\beta_{\Omega}$ are shown in panels (B,D).  
While  $\beta_{\Omega}$ has the correct  overall form, the deviations from the 
canonical curves are clearly visible. 

The extent of the deviation is shown quantitatively in Figure \ref{fig:S_omega}(E) 
by plotting the root-mean-square (RMS) distance between the canonical and 
resultant temperature, versus Hilbert space dimension $D$, for an $L$-site spin 
chain ($D=2^L$).  The deviation decreases very slowly --- it would take very large 
system sizes for the two temperatures to be considered `close'.

We now discuss two ways of understanding this finite size deviation.
First, using Eq.\ \eqref{eq:def-OmegaEntropy} for the entropy means omitting the 
$\ln\dE$ term from the definition, Eq.\ \eqref{eq:entropy_from_dos}, which can 
be written as 
\begin{equation}\label{eq:S_Somega}
S = S_{\Omega} + k_B\ln\dE . 
\end{equation}
Since we avoided making an explicit choice for $\dE$, it is not obvious what the effect of dropping
the second term is, but we can analyze different cases:
\begin{itemize}
\item If we consider $\dE$ to be energy-independent, then $S_{\Omega}$ will lead to the same
  temperature as obtained from the full microcanonical entropy $S$.  However, we know from Section
  \ref{sec:const_dE} that $S$ obtained using an energy-independent $\dE$ leads to considerable
  finite-size deviations in the temperature.
\item On the other hand, if $\dE$ were to be energy-dependent, e.g., if it were designed to cancel
  the sub-leading deviations from the canonical ensemble as in Section \ref{sec:energy_dependent_dE},
  then $S_{\Omega}'(E)$ will differ from $S'(E)$, and we would again get finite-size deviations.
\end{itemize}
Thus, it appears that $S_{\Omega}'(E)$ can be expected to deviate from $\beta_c$ in either case.  



Second, we note that Eq.\ \eqref{eq:def-OmegaEntropy} applies the logarithm to a dimensionful
quantity, which strictly speaking is not allowed.  For consistency, one needs to multiply the
argument of the logarithm in Eq.~\eqref{eq:def-OmegaEntropy} by a quantity having the dimensions of
energy; let us call this quantity $\epsilon$. Then Eq.\ \eqref{eq:def-OmegaEntropy} should really
have the form
\begin{equation}
S_{\Omega}=\ln\left[g(E)\epsilon\right] = \ln\left[\Omega'(E)\epsilon\right] . 
\end{equation}
One can then carry out the same saddle point approximation as previously performed (subsection
\ref{sec:saddlepoint}) with the original definition of $S$, leading to
\begin{equation}\label{eq:omega_saddle}
\frac{S_{\Omega}}{k_B} \approx \beta_c E- \beta_c F(\beta_c)) + \ln\epsilon -\ln\sqrt{2\pi k_{B}T_{c}^{2}C_{c}}.
\end{equation}
Unless $\epsilon$ is carefully chosen (as we did for $\dE$ in Section
\ref{sec:energy_dependent_dE}), the last term will lead to finite-size deviations.  The procedure in
this section avoids specifying $\dE$ and ignores the need for the quantity $\epsilon$.  Thus, one
would expect deviations due to the $-\ln\sqrt{T_{c}^{2}C_{c}}$ term, essentially of the same type as
that encountered in \ref{sec:const_dE} when using an explicit energy-independent value of $\dE$.

\subsection{Using the integrated D.O.S. with an energy dependent window} \label{sec:omega_eps}

Now, rather than neglect $\dE$ in Eq.~\eqref{eq:entropy_from_dos}, we retain the term, i.e., we consider the entropy in terms of the integrated d.o.s.\ as 
\begin{equation}\label{eq:def_OmegaEntropy_full}
	S(E) = k_B \ln(g(E)\dE) = k_B \ln(\pd{\Omega(E)}{E}\dE).
\end{equation}

We have a smooth approximation to $\Omega'(E)$, and as proposed in Section 
\ref{sec:energy_dependent_dE}, we could choose an energy-dependent $\dE(E) \propto \sqrt{T_c^2C_c}$ to 
account for the finite size deviations. 
Here, the constants of proportionality will only result in a shift in entropy, as we are not counting eigenstates in an energy window of width $\dE(E)$ around $E$. In that case, we had to choose a constant $\alpha$, such that the window width was not comparable to the bandwidth of the spectrum. 

Here, rather than count the number of eigenstates, we use the derivative of our smoothed approximation to the integrated d.o.s.\ to compute $g(E)$. The proportionality constants in this case are arbitrary, thus our entropy can now be written as
\begin{equation}\label{eq:ent_omegaANDeps}
	S_{\Omega}(E) = \ln\left[\Omega'(E)\sqrt{T_c^2 C_c}\right] .
\end{equation}
Again, we use $\Omega'(E) = \frac{D}{2}\sech^2(f(E))f'(E)$. Determining the canonical temperature $T_c$ and heat capacity $C_c$ numerically has been previously discussed (subsection \ref{sec:fin_sys_res}). 
Thus we can calculate the entropy \eqref{eq:ent_omegaANDeps} and by fitting a polynomial to the resulting curve, we can derive the inverse temperature.

\begin{figure}
	\centering
	\hspace{0.10\linewidth}
	\includegraphics[width=0.3\linewidth]{Lattice_tikzit.pdf}
	\hspace{0.15\linewidth}
	\includegraphics[width=0.3\linewidth]{Square_tikzit.pdf}\\
	\includegraphics[width=0.99\linewidth]{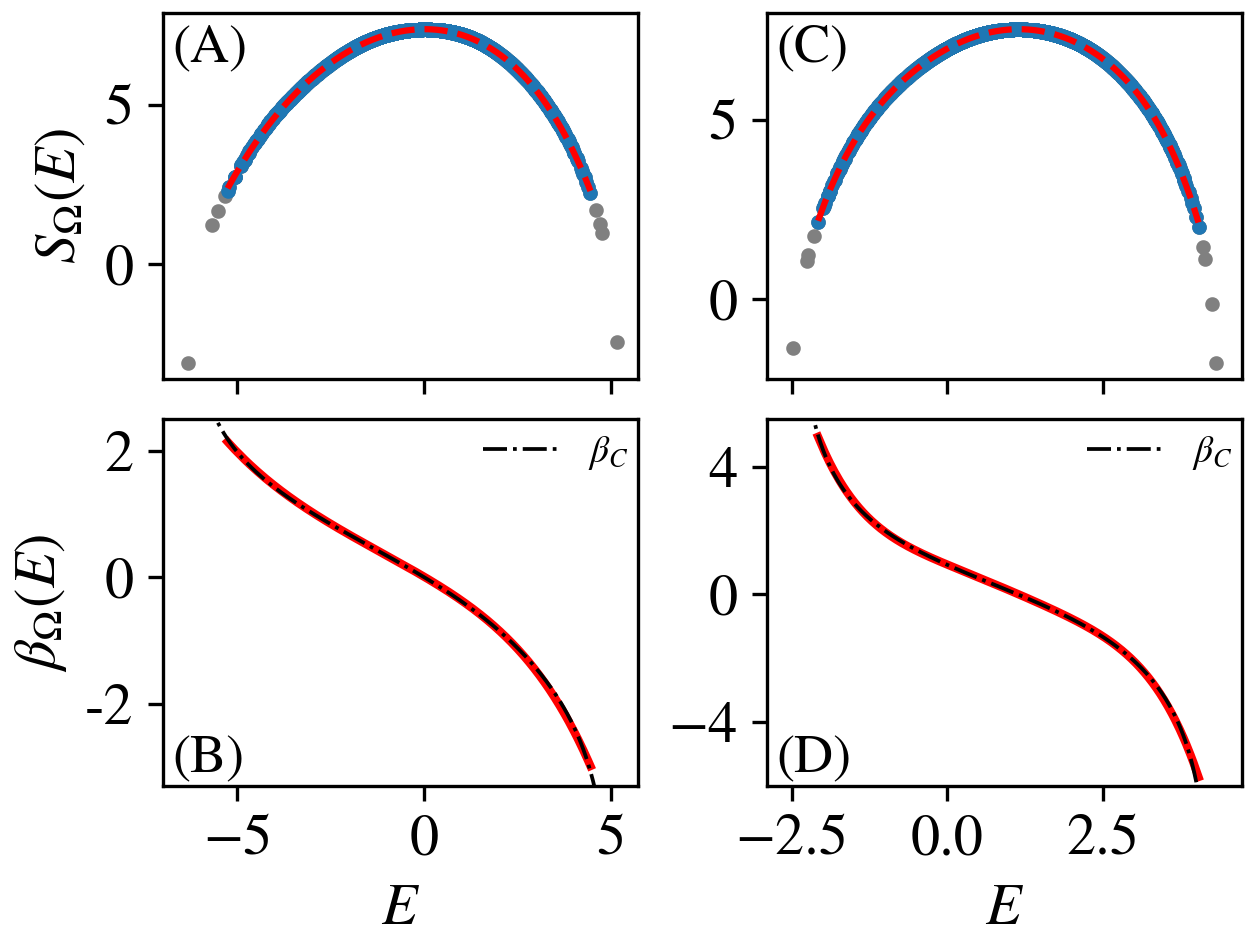}
	\caption{  \label{fig:S_omega_eps}
		Entropy and temperature using a continuous approximation to the cumulative density of states 
		and an energy dependent $\dE(E) = \sqrt{T_c^2 C_c}$. 
		\textbf{(A,C)} $S_{\Omega}(E) = \ln(\Omega'(E)\dE(E))$. 
		\textbf{(B,D)} Resultant inverse temperature $\beta_{\Omega}(E) = \partial_E S_{\Omega}(E)$ compared with the canonical temperature $\beta_C$. 
		\textbf{(A,B)} Staggered field \textit{XXZ} chain, Eq.\ \eqref{eq:Ham_stag},  with $L=12$, $J=1$, $\Delta=0.95$,
		$h_z = h_x = 0.5$.  
		\textbf{(C,D)} $5\times4$ square \textit{XXZ} lattice, Eq.\ \eqref{eq:Ham_square},  with $N=4$.
	}
\end{figure}

Examples of the computed entropy $S_{\Omega}(E)$ are shown in Figure \ref{fig:S_omega_eps}(A,C). 
The numerical results shown are for a 1D spin chain and a square lattice, both 
with nearest neighbor \textit{XXZ}-like connections and open boundary conditions. 
The resultant inverse temperatures $\beta_{\Omega}(E)$ are shown in panels (B,D).  
We see that $\beta_{\Omega}(E)$ aligns very well with the canonical temperature $\beta_{c}$.
A pleasing aspect of this method is that it requires no fine-tuning of constants, as they only 
result in a shift of entropy.

\section{Concluding Discussion \& Context}  \label{sec:concl}

The equivalence of microcanonical and canonical ensembles emerges in the infinite-size limit.
Since statistical mechanics in isolated systems is being intensively
discussed through finite-size examples obtained by numerical diagonalization, it is important to
understand deviations from ensemble equivalence in systems of such sizes.  In this work, we
contribute to this question by investigating and comparing various ways of computing the
microcanonical entropy, and then comparing the resultant temperatures (via Eq.\ \eqref{eq:entropy_temp})
to the canonical temperature (Eq.\ \eqref{eq:canonical_energy}).

The microcanonical entropy $S(E)$ is defined as in Eqs.\ \eqref{eq:entropy_def} and \eqref{eq:dos}.
Inspired by the discussion of Ref.~\cite{Gurarie_Equivalence_AJoP2007}, we explored four ways of
calculating $S(E)$ numerically: in terms of a constant-width energy window (Section
\ref{sec:const_dE}), using an energy-dependent window with the energy-dependence designed to cancel
sub-leading terms (Section \ref{sec:energy_dependent_dE}), and using the integrated density of states, first with an approximation that avoids the
energy window altogether, and alternatively, in combination with an energy-dependent window (Section \ref{sec:Omega}).

We have demonstrated that counting eigenstates using the energy-dependent window,
$\dE(E) = \alpha^{-1}\sqrt{2\pi k_B T_{c}^{2}C_{c}}$, works extremely well for the sizes under
consideration.  The use of this energy-dependent window was suggested by Ref.\
\cite{Gurarie_Equivalence_AJoP2007} with $\alpha=1$, designed to exactly cancel the sub-leading terms
in Eq.\ \eqref{eq:entropy_final}.  We have shown (Figure \ref{fig:S_eps}) that, for the sizes under
question, a larger $\alpha$ is required, because for $\alpha=1$ the windows exceed the
energy scale of variation of the density of states.  A consequence of this result is that, since we
have no further criterion for fixing $\alpha$, the microcanonical entropy is only defined up to an
arbitrary additive constant. The constant is sub-extensive (and thus unimportant in the
infinite-size limit), and an additive constant does not affect the temperature.  
In fact, a pleasing aspect of the suggestion of Ref.\ \cite{Gurarie_Equivalence_AJoP2007} was 
a reasonable criterion for defining entropy without such an arbitrary constant.  
For the size ranges under consideration in this work, the $\alpha=1$ prescription is not usable, 
so we are forced to accept an arbitrary shift.

However, using the entropy $S_{\Omega}(E)$ in Section \ref{sec:omega_eps}, combining the use of an energy-dependent $\dE(E)$ with an approximation to the d.o.s.\ constructed from the integrated d.o.s.\ $\Omega$, we observed excellent agreement between the resulting temperature and the canonical temperature. Here, the constants in $\dE(E)$ do not affect the resulting temperature, as they simply shift the entropy; this method is thus free from any fine-tuning of constants.

The other two procedures (counting eigenstates in an energy-independent $\dE$ and $S\approx S_\Omega$) 
both lead to very noticeable finite-size deviations. The deviations in the constant-$\dE$ case are 
expected from the sub-leading corrections.  For the procedure in Section \ref{sec:Omega} using 
$S\approx k_B\ln(\Omega'(E))$, we have argued that the deviations are essentially of the 
same type as that in the constant-$\dE$ case.


Numerical calculations of microcanonical entropy and/or temperature have appeared in the recent
literature \cite{SantosPolkovnikovRigol_Entropy_PRL2011,Kourehpaz_TypicalityChaos_ent2022,
  RussomannoFavaHeyl_Entropy_PRB2021, SilvaGoold_Temperature_PRA2022}.  Refs.\
\cite{Kourehpaz_TypicalityChaos_ent2022, SilvaGoold_Temperature_PRA2022} have used the
$S\approx S_\Omega$ approximation.  The resulting temperature shown in Ref.\
\cite{Kourehpaz_TypicalityChaos_ent2022} (Figure 4) appears to have similar deviations from the
canonical temperature as we have presented and analyzed in Section \ref{sec:Omega}.  The density of
states is approximated in Ref.\ \cite{Kourehpaz_TypicalityChaos_ent2022} via a polynomial fit to
the cumulative spectral function and in Ref.\ \cite{SilvaGoold_Temperature_PRA2022} via the kernel
polynomial method \cite{WeisseFehske_KernelMethod_RMP2006,CiracBanuls_PRL2020_spectralanalysis}.  In
Ref.~\cite{SantosPolkovnikovRigol_Entropy_PRL2011}, the microcanonical entropy has been calculated
with $\dE$ ``determined by the energy uncertainty'' --- this is presumably equivalent to the
energy-dependent window we explored in Section \ref{sec:energy_dependent_dE}.

For finite quantum systems, the entanglement entropy of a subsystem is often discussed as
representing the thermal entropy \cite{Deutsch_Entropy_IOP2010, Deutsch_Sharma_PRE2013,
  GarrisonGrover_SingleEstate_PRX2018, LuGrover_RenyiEntropy_PRE2019, SekiYunoki_thermal_PRR2020}.
The reason is that, in a chaotic (thermalizing) quantum system, it is expected that the reduced
density matrix of subsystems smaller than half the system should resemble thermal density matrices
\cite{LindenPopescu_ThermEq_PRE2009,Muller2015, Greinergroup_thermalization_Science2016,
  Dymarsky_Lashkari_Liu_PRE2018, GarrisonGrover_SingleEstate_PRX2018, SekiYunoki_thermal_PRR2020,
  CBurkeNakerstHaque_Temperature_PRE2023}.  It is amusing to note that the entanglement entropy is
obtained from eigenstates, whereas the entropies and temperatures studied in this work are derived from eigenenergies.  Ref.\ \cite{GarrisonGrover_SingleEstate_PRX2018} describes the finite-size behavior of the deviation of the entanglement entropy from the canonical entropy for a particular
spin chain.  The behavior of the temperature derived using Eq.\ \eqref{eq:entropy_temp} from the
entanglement entropy (interpreted as the entropy) would be interesting to examine in future work.

It has been argued that entropy for finite-size systems should be defined as $S\sim \ln\Omega(E)$
\cite{PearsonEtAl_GibbsEntropy_PRA1985, TalknerHanggiMorillo_MicrocanFluctuation_PRE2008,
  DunkelHilbert_NegativeT_Nat2014, HilbertHanggi_ThermoIsolated_PRE2014,
  Campisi_MicroEntropy_PRE2015, Hanggi_Temperature_RoyalSoc2016}, instead of the more common
$S\sim \ln\Omega'(E)$ examined here.   This approach avoids the appearance of negative
temperatures even in systems with finite Hilbert space dimension.  
It may be interesting to ask how the deviations from ensemble equivalence behave under this
alternate definition. 
   

\section*{Acknowledgments}
\noindent\textit{Acknowledgments.-} PCB thanks Maynooth University (National University of Ireland,
Maynooth) for funding provided via the John \& Pat Hume Scholarship.  This work was in part
supported by the Deutsche Forschungsgemeinschaft under grants SFB 1143 (project-id 247310070).  The
authors acknowledge the Irish Centre for High-End Computing (ICHEC) for the provision of
computational facilities.  The authors are grateful for useful discussions with Tarun Grover, Goran
Nakerst, and Paul Watts.

\normalem
%


\end{document}